\documentclass[reprint,
pre,
notitlepage]{revtex4-1}

\usepackage{graphicx} 
\usepackage{float}
\usepackage{array} 
\usepackage{paralist} 
\usepackage{subfigure} 

\usepackage{fancyhdr} 
\usepackage{threeparttablex}
\usepackage{amsmath}
\usepackage{tcolorbox}
\usepackage{tikz-inet}	
	\usetikzlibrary{graphs,trees,shapes,snakes,arrows,circuits.logic.US}
	\usepackage{circuitikz}
	\usetikzlibrary{backgrounds,fit,decorations.pathreplacing} 
	\tikzstyle{branch}=[fill,shape=circle,minimum size=3pt,inner sep=0pt]
	
	\usepackage{xcolor}
\newcommand{\ket}[1]{\ensuremath{\left|#1\right\rangle}} 
\newcommand{\bra}[1]{\ensuremath{\left\langle#1\right|}} 

\begin{document}
\title{Quantum IsoRank: Efficient Alignment of Multiple Protein-Protein Interaction Networks} 
 \author{Anmer~Daskin}\email{email:anmerdaskin@yahoo.com}
\affiliation{Department of Computer Engineering, Istanbul Medeniyet University, Uskudar, Istanbul, Turkey}
\begin{abstract}

Comparative analyses of protein-protein interaction networks play important roles in the understanding of biological processes. However, the growing enormity  of available data on the networks  becomes a computational challenge for the conventional alignment algorithms. Quantum algorithms generally provide greater efficiency over their classical counterparts in solving various problems.
 One of such algorithms is the quantum phase estimation algorithm which generates the principal eigenvector of a stochastic matrix with probability one. 
 
 Using the quantum phase estimation algorithm,  we introduce a quantum computing approach for the alignment of protein-protein interaction networks by following the classical algorithm IsoRank which uses the principal eigenvector of the stochastic matrix representing the Kronecker product of the normalized adjacency matrices of networks for the pairwise alignment.
We also present a  greedy quantum measurement scheme to efficiently procure the alignment from the output state of the phase estimation algorithm where the eigenvector is encoded as the amplitudes of this state. 
The complexity of the quantum approach outperforms the classical running time.
\end{abstract}

\maketitle
\markboth{Quantum IsoRank}{Quantum IsoRank}
\section{Introduction}
Comparative analyses of protein-protein interaction (PPI) networks play important roles in the understanding of biological processes. 
Alignments of PPI networks drawn from different species provide invaluable information to catalog conserved network regions and identify functional similarities across species. 
Using different formulations, many network alignment algorithms have been proposed such as  the ones 
in  \cite{KLau2009, Kelley2004, Koyuturk2006pairwise,Li2007alignment,Zaslavskiy15062009} and  others \cite{Clark2014comparison}.  However, the growing enormity  of available data on the networks  raises computational challenges for  implementations of these algorithms. 
In analogous to ranking algorithms, e.g. PageRank \cite{PageRank}, one of the global alignment algorithms, IsoRank \cite{Singh2007}, uses the intuition that the score of aligning two nodes  should depend on the alignment of their neighbours in the global alignment of two PPI networks. 
It formulates the network alignment  as an eigenvector problem where coefficients of the principal eigenvector of a stochastic matrix, the stationary state, represent the functional similarity scores between pairs of the nodes.
Using a greedy algorithm,   it then generates the network alignment from  the eigenvector. 
In the case of multiple networks, IsoRank is applied to every pair of the networks and a global alignment is retrieved from the pairwise network alignment results \cite{IsorankN2009}. 
While IsoRank is one of the successful algorithms and fast enough to handle the alignment of large sparse graphs \cite{Wang2009}, 
the exponential scaling of its running time  with the number of networks still impedes applications of the algorithm to multiple networks (the running time of the algorithm is $O(E^m )$ where $E$ is the
number of edges in a network and $m$ is the number of networks) \cite{Noble2008}.

In \cite{Daskin2014mna}, it has been discussed that aligning networks on quantum computers may provide  greater efficiency. 
In particular, it is shown that the quantum phase estimation algorithm can be used for stochastic matrices to find their principal eigenvector with the success probability one.  
However,  it has  not been shown how to  procure a solution from the quantum state representing the principal eigenvector. 
It is also not shown how to simulate ranking matrices which are not Hermitian in most of the cases in the phase estimation algorithm.

Here, we first show explicitly how  the eigenvector of a stochastic matrix can be obtained  by using the quantum phase estimation algorithm by following mostly Ref.\cite{Daskin2014mna}. 
To construct the alignment from the final quantum state, we then give a greedy quantum algorithm  based on a quantum measurement scheme.
 In addition, we present an approximate simulation approach for non Hermitian ranking matrices and present simple examples with numerical results which can be  experimented on quantum computers (The capacities of the current quantum computers are very limited. 
 Thus, they can only run for small sized problems.). In the end,  the complexity analysis  shows that the multiple network alignment by this quantum approach  requires 
 $O\Big(m\times poly\big(log(|V|)\big)\Big)$ computational running time for sparse matrices while $O\big(m|V|^2\big)$ for dense matrices. This is an exponential speedup over the classical running time.

\section{Classical IsoRank}

A PPI network  is generally represented as an undirected graph,
$G(V, E)$ where the set of nodes $V$ represents the set of proteins and the set of edges $(v_i, v_j) \in E$
describes interactions between proteins $v_i$ and $v_j$. 
To observe conserved similarities across species, PPI networks for different species are comparatively analyzed by maximizing an objective function to indicate correspondences between the nodes. One of the most common strategies to observe the similarities between networks is the network alignment. Network alignment is the process to globally compare two or more networks to identify the regions where these networks are similar and dissimilar. 
In general terms, network alignment can also be  considered as a more general variant of the subgraph isomorphism problem  which is an NP-complete problem  of determining  whether a finite larger graph contains a smaller graph  as an exact subgraph \cite{Cook1971}.  However, in the network alignment, it is important to find the most similar and dissimilar parts between two graphs even if the larger graph  does not contain the smaller one as an exact subgraph.\cite{Sharan2006modeling} 

Singh et al.\cite{Singh2007} have presented a global alignment algorithm, viz. IsoRank, by using the intuition that the score of aligning two nodes  should depend on the alignment of their neighbors. This intuition is formulated as follows:
\begin{equation}
	  \label{Eq:sing1}
	  R_{ij}= \displaystyle\sum_{u\in N(i)}\sum_{v\in N(j)}\frac{1}{|N(u)||N(v)|}R_{uv},
\end{equation}
where, $N(a)$ is the set of neighbors of the node $a$; $|N(a)|$ is the size of this set;
$V_1$ and $V_2$ are the set of the nodes for the networks $G_1$ and $G_2$; and $i\in V_1$ and
$j\in V_2$. $R$ defines the functional similarity matrix whose stationary state is
used to find the solution for the alignment problem.  Eq.(\ref{Eq:sing1}) can be rewritten also in matrix form as:

\begin{equation}
	  \label{Eq:IsoRank1}
	  \begin{split}
	  R&=AR,\\
	  A[i,j][u,v]&=
	  \Big\{\begin{matrix}
	  \frac{1}{|N(u)||N(v)|} & 
	  \text{if $(i,u) \in E_1$ and $(j,v) \in E_2$,} \\
	  0 & \text{otherwise.}
\end{matrix}	   
\end{split}
	  \end{equation}
In the above equation, $A$ is a $|V_1||V_2|$x$|V_1||V_2|$ stochastic matrix and can be defined from the Kronecker product of the column-wise normalized adjacency matrices of the input graphs:
$A=A_1\otimes A_2$, where $A_i$ is the normalized adjacency
matrix for the graph $G_i$ and also a stochastic matrix. 
Eq.(\ref{Eq:IsoRank1}) describes an eigenvalue problem where the principal eigenvector corresponding to the eigenvalue one of the matrix $A$ is the stationary distribution of the
random walk on the Kronecker product graph.
On classical computers, this equation can be solved 
by using different iterative methods such as the power iteration.  

\section{Quantum Phase Estimation Algorithm for Network Alignment}
In this section, we  briefly explain the main intuition of the quantum phase estimation algorithm (for an unfamiliar reader, we recommend Ref.\cite{Nielsen2010quantum}) and then  show how the stochastic matrices can be used in the phase estimation algorithm by following mostly Ref.\cite{Daskin2014mna}. Then, we discuss how to simulate a non-Hermitian stochastic matrix on quantum computers by using the closest Hermitian matrix. In the final subsection, we show how to incorporate some other information used in the PPI networks.

\subsection{Quantum Phase Estimation Algorithm}
For a given approximate eigenvector encoded as the amplitudes of the quantum state $\ket{\mu_j}$ 
and the eigenvalue equation  $U\ket{\mu_j}=e^{i2\pi\phi_j}\ket{\mu_j}$; the phase estimation algorithm (PEA) \cite{Kitaev1995} tries to find the phase $\phi_j$ in this equation. 
PEA mainly requires two quantum registers, \ket{reg_1} and \ket{reg_2}, consisting of sufficient number of qubits to hold the eigenvector and the phase, respectively. 
 In the initial setting, $\ket{reg1}$ is set to zero state and $\ket{reg_2}$ is assigned to hold a vector which is the best known approximation of \ket{\mu_j}.
 With the help of quantum Fourier transform 
 and the sequential controlled unitary operations, \ket{reg_1} becomes holding the Fourier transform of the phase. 
Then, the application of the inverse  quantum Fourier transform turns   $\ket{reg1}$  into the binary value of the phase: $\ket{reg1}=\ket{\phi_j}$.  
 Consequently,  the value of the phase is obtained by measuring \ket{reg1} in the standard basis.  
Here, if the unitary operator $U$ is the time evolution operator of a Hermitian matrix $H$, $U=e^{i2\pi H}$, 
 then one also obtains the eigenvalue of $H$.
 
\subsection{Application to Stochastic Matrices}
The success of the phase estimation algorithm is directly related to the closeness of the input vector  to the actual eigenvector. 
This can be defined by the dot product. 
The dot product of an equal superposition state and a vector is the normalized sum of the vector elements. 
On the other hand, the eigenvectors of a stochastic matrix has the property that the sum of the vector elements is one  for the principal eigenvector and zero for the rest of the eigenvectors. 
In \cite{Daskin2014mna}, it has been showed that when PEA is given an equal superposition input state,   it then finds the principal eigenvector of $U$  and so the principal eigenvector of the stochastic matrix $H$ with the success probability equal to one. 

In our case, we find the principal eigenvector of the matrix $A$ which is the Kronecker product of the column-wise normalized  adjacency matrices (normalized in the sense to have stochastic matrices) for the input networks: i.e., $A=A_1\otimes A_2 \otimes \dots \otimes A_m$. Because of the Kronecker product, $A$ defines a separable system and so quantum circuits for each $A_i$ can be constructed separately. 
This  eases the difficulty of finding quantum circuits  for the simulation. 
However, in general, quantum computing is based on  unitary gates associated with time evolution operators of Hermitian quantum systems. This dictates the stochastic matrix used in the algorithm to be Hermitian, in which case the principal eigenvector is already known to be a vector of all ones. In the following subsection, we shall describe how to approach non-Hermitian matrices.   

 \subsection{Simulation of  Non-Hermitian Operators}
A matrix $A$ is called positive if the matrix elements 
$A_{ij}> 0$ and non-negative if $A_{ij}\geq 0$. 
It is normal if $A^\dagger A-AA^\dagger=0$, where $A^\dagger$ describes the conjugate transpose of $A$. 
 Any matrix $A$ can be decomposed into a Hermitian and a skew-Hermitian matrices as:
 \begin{equation}
 \label{EqHSH}
 A=H+S=\frac{1}{2}(A+A^\dagger)+\frac{1}{2}(A-A^\dagger),
 \end{equation}
 where  $H=\frac{1}{2}(A+A^\dagger)$ and 
 $S=\frac{1}{2}(A-A^\dagger)$ define the nearest Hermitian and 
 skew-Hermitian matrices to $A$, respectively \cite{ClosestHermitian1975}. 
 The eigenvalues of $H$ are all real and the eigenvalues of $S$ have only imaginary parts. Moreover, 
 when $A$ is a normal matrix, there are a few additional useful properties:
 \begin{itemize}
   \item $H$ and $S$ commute: $[H,S]=HS-HS=0$. 
 \item Since $AA^\dagger=A^\dagger A$, $H$ and $S$ have the same eigenvectors.
 \item  The imaginary part of the eigenvalues of $A$ are equal to the eigenvalues of $S$, and the real parts are equal to the eigenvalues of $H$. 
\end{itemize}  
Because of the last property, one can simulate normal matrices and their corresponding non-Hermitian operators on quantum computers by using two separate registers to obtain the imaginary  and the real parts of the eigenvalue individually. In that case, one uses two unitary operators $U_1=e^{iH}$ and $U_2=e^{S}$  for the simulation. (Note that $U_2$ is a unitary matrix because the exponential of a skew symmetric matrix is a unitary matrix.).
However, if a stochastic matrix is normal, it turns out that it is also doubly stochastic: i.e., its left and right principal eigenvectors are known to be a vector of all ones with the eigenvalue one. 
Therefore, instead of an approximate normal matrix, we shall use the closest Hermitian matrix $H=\frac{1}{2}(A+A^\dagger)$ in our simulations. $H$ defines a nonnegative matrix. If it is also irreducible, then there is a well known theorem stating that $H$ has a dominant positive eigenvalue equal to the spectral radius, $\rho(H)$, of matrix $H$ and corresponds to a real nonnegative eigenvector unique up to a scalar multiple \cite{Berman1994}. The value of $\rho(H)$ is bounded as:
$min_j\sum_{i}h_{ij} \leq \rho(H) \leq max_j \sum_{i}h_{ij}$, where $h_{ij}$ are the matrix elements of $H$. Using the stochastic property of $A$,  we also find  $\rho(H) > 0.5$. 
PEA outputs $j$th eigenvector with the probability: 
 \begin{equation}
\left|\frac{1}{\sqrt{N}}\sum_i^N\bra{\textbf{i}}\ket{\mu_j}\right|^2.
\end{equation}
This is the normalized sum of the coefficients of the $j$th eigenvector and is equal to one if $\ket\mu_j$ is the principal eigenvector and zero for the rest of the eigenvectors. Therefore, the success probability of PEA for stochastic matrices is one.
Moreover, since $H$ is the best Hermitian approximation to $A$, the principal eigenpair of $H$ close to the ones of $A$. Hence, we expect the sums of the coefficients of the eigenvectors of $H$ to be very similar to the ones of the eigenvectors of $A$ and so the success probability to be very close to one.

\subsection{Incorporation of Other Information}
As done in IsoRank \cite{Singh2007}, one can include further information, e.g. BLAST scores, into the quantum model as well in  the following form:
\begin{equation}
\tilde{A}=H+B,
\end{equation}
where $B$ and $H$ are to be assumed to commute: $BH-HB=0$. 
Therefore, the time evolution can be written as:
 \begin{equation}
e^{i\tilde{A}}=e^{iH+B}=e^{iH}e^{iB}
\end{equation}
Note that the above equation does not change the sparsity of the matrix $A$; hence, it is still sparse and the evolution operator and the corresponding quantum circuit require polynomial time for the implementation \cite{Berry2007sparse} (see Sec.\ref{Sec:Complexity} for the complexity analysis).

\begin{tcolorbox}
\subsection{Summary of the Method}
\begin{itemize}
\item Prepare to quantum registers as: \ket{reg_1}=0 and $\ket{reg_2}=\frac{1}{\sqrt{N}}\sum_i^N\ket{\textbf{i}}$ defining  an equal superposition state.
\item Find $H=\frac{A+A^\dagger}{2}$ as the closest Hermitian approximation to the stochastic matrix $A$.
\item Find the quantum circuit simulating $H$ by using one of the methods described in \cite{Berry2007sparse,Childs2011}.  The number of gates in the resulted circuit is polynomially bounded w.r.t. the number of qubits for the sparse matrices.
\item Apply the phase estimation algorithm with the above initial settings.
\item For stochastic matrices; the PEA results the correct answers with probability one: the probability of seeing $j$th eigenvector is given by the normalized square of the sum of its coefficients:
 \begin{equation}
\left|\frac{1}{\sqrt{N}}\sum_i^N\bra{\textbf{i}}\ket{\mu_j}\right|^2.
\end{equation}
 Since the sum of the coefficients are one for the principal eigenvector and zero for the rest, PEA results the principal eigenvector with probability one.
\item For the closest Hermitian matrix $H$; the probability is not one. However, it is still expected to be very close to one since $H$ is a nonnegative matrix with a principal eigenvalue and a positive eigenvector (this is due to the Perron-Frobenious theorem) close to the eigenvector of the original stochastic matrix. 
\end{itemize}
\end{tcolorbox}
\vspace{1in}
\section{Extracting Node Mappings From a Quantum State}
\subsection{Matching for a Pair of Networks}
Generating a discrete solution from the final quantum state is known to be just solving a maximum weight matching problem. However, since fully obtaining a quantum state requires exponential time complexity \cite{Nielsen2010quantum}, we cannot apply classical matching algorithms directly. 
Hence, engineering the order of the quantum registers in the measurement, we describe the following greedy strategy for the alignment: 
Consider the eigenvector consists of two registers as
$\ket{\textbf{x}}\ket{\textbf{y}}$, where each register represents a network:
\begin{enumerate}

\item Apply a conditional measurement: when the first register is $x_i$, measure the second register. For each measured $x_i$, this generates a $y_j$; therefore,  $x_i$s are matched to $y_j$s. Note that the measurement outcome is determined by the conditional probability (The probability is conditional when additional information such as BLAST scores is included. Otherwise, it is disjoint.): the probability of measuring $y_j$ in the second register while the first register is $x_i$. 
\item If there are duplicities and still unmatched nodes;
 then apply a second type of conditional measurement: if the second register $y_j$, measure the first register. This time for each $y_j$ an $x_i$ is obtained. 
 
 This step is only useful when additional information is incorporated ($A+B$ is used) because the system is otherwise separable as $A=A_1\otimes A_2$ and the conditional measurement outcome will be the same as the outcome obtained in the first step.
\item We combine these two different measurement outcomes obtained in the first and the second steps and begin matching from nodes whose scores are the highest. 
\item If there are still unmatched nodes,  statistical information about the other possibilities  obtained during the measurement is used to match the remaining nodes.
\end{enumerate}

Since the main intuition of the algorithm gives a higher similarity score to the nodes whose neighbours have high scores; in the matching,  the neighbours of the first matched nodes are given priority so as to generate a solution which is  also connected. 
In the end,  we choose the largest connected component as the best solution to the alignment.

\subsection{Matching for Multiple Networks}
\label{sec:MatchingMN}
In the case of multiple networks, as done in \cite{IsorankN2009} an obvious approach is to align each possible pair of networks and derive a common solution for the multiple networks from these pairwise  alignments. 
However, the complexity of this approach grows exponentially with the number of networks. 
Here, we shall follow a different approach which can be generalized easily to any number of networks: 
Assume we have three networks $G_1$, $G_2$, and $G_3$ with nodes $\{a_1\dots a_{m_1}\}$, $\{b_1\dots b_{m_2}\}$ and $\{c_1\dots c_{m_3}\}$, respectively.
 Let also $\ket{reg_{G_1}}$, $\ket{reg_{G_2}}$, and $\ket{reg_{G_3}}$ represent the networks $G_1$, $G_2$, and $G_3$, respectively:
\begin{itemize}
\item Measuring \ket{reg_{G_1}} alone and  \ket{reg_{G_2}} and \ket{reg_{G_3}} together, we draw the conditional probabilities to see one of the nodes, $a_i$, in \ket{reg_{G_1}} and $b_jc_k$ in \ket{reg_{G_1}}\ket{reg_{G_3}}.
In other words, when \ket{reg_{G_1}} is $a_i$, the probability to see $b_jc_k$ in \ket{reg_{G_1}}\ket{reg_{G_3}} is obtained. 
This results in a matching of $a_i- b_jc_k$. 
Note that in the real implementation on a quantum computer, one just assigns the nodes as a result of the measurement outcome. 
\begin{itemize}
\item We match nodes initiating from  the largest probability, or the most commonly seen measurement outcome.
\item As done in the case of two networks, the priority given to the neighbours of the first matched nodes.
\end{itemize}
\item  If the measurement is no longer provide valuable information to match further nodes, then we measure \ket{reg_{G_2}} alone and  \ket{reg_{G_1}} and \ket{reg_{G_3}} together.  This gives the  probabilities to see one of the nodes, $b_j$, in \ket{reg_{G_2}} and $a_ic_k$ in \ket{reg_{G_1}}\ket{reg_{G_3}}. Then we combine this measurement result with the previous measurement result to match the unmatched nodes.
\end{itemize}  

One can also  go further and draw probabilities for $c_k-a_ib_j$ and combine them with the previous results. 
However, while different measurement settings increase the statistical confidence in the measurement results, it increases the complexity of the algorithm. As noted before, if \ket{reg_2} is separable at the beginning: i.e., $\ket{reg_2}= \ket{reg_{G_1}} \otimes \ket{reg_{G_2}}\otimes \dots \otimes \ket{reg_{G_m}}$, then the other measurements also produce the same result as the first measurement setting, $G_{1i}-G_{2j}\dots G_{mk}$. 
Thus, only the first measurement setting is used to conclude the matching of the nodes.

\begin{figure*}[ht]
\centering
\subfigure[ $G_1$]{
\includegraphics[scale=0.5]{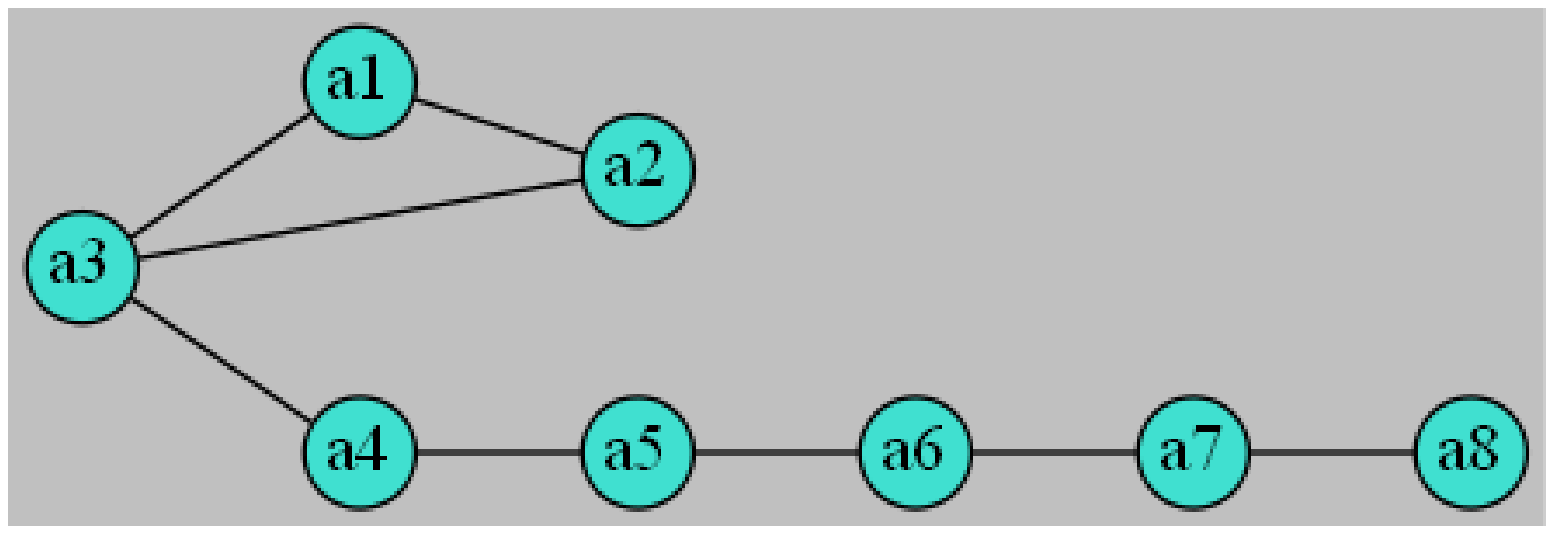}}
\subfigure[ $G_2$]{
\includegraphics[scale=0.5]{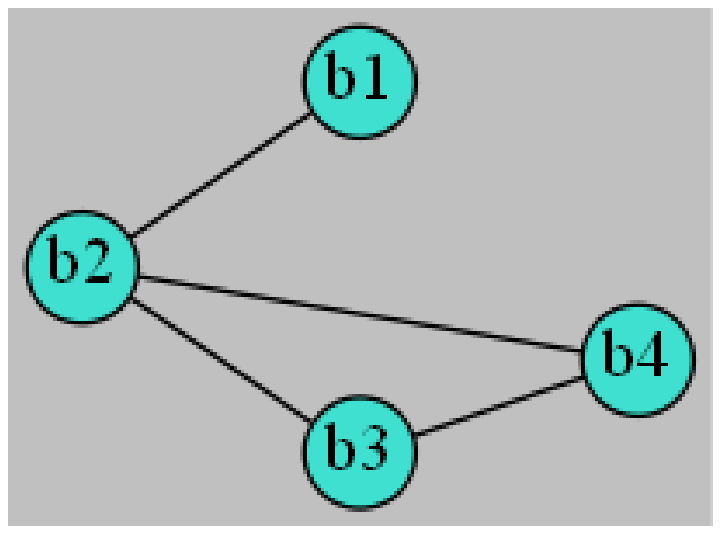}}
 \subfigure[ $G_3$]{
\includegraphics[scale=0.5]{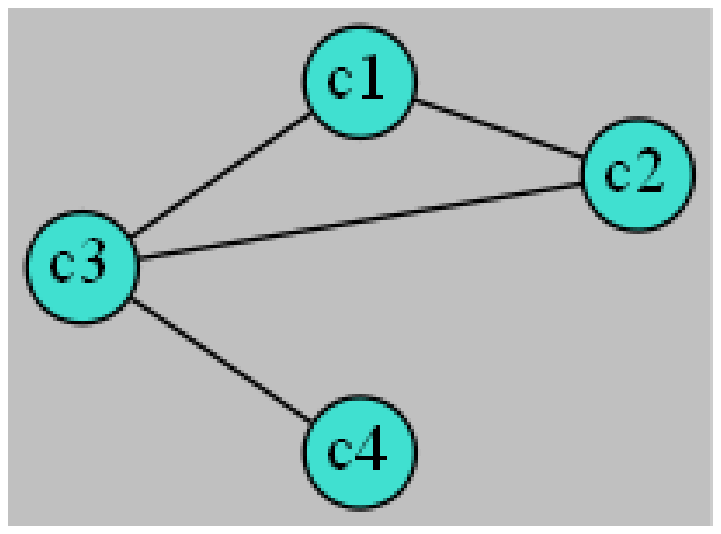}}
\caption{\label{fig:networks} Example Networks }
\end{figure*}

\section{Examples}

Because of the computational difficulty in simulating quantum computers on classical computers 
(The complexity grows exponentially with the number of qubits involved in the simulation. PEA requires also sufficient number of qubits on the first register to hold eigenvalue.),  
here we only aim to show the capability of the algorithm when a fully functional quantum computer is available. 
Therefore, we shall use merely trivial examples given 
in Fig.\ref{fig:networks} for which one can see optimal solutions easily. 
For the large non-trivial  networks, since both IsoRank and the quantum approach are based on the principal eigenvector; we expect the success of the algorithm to be similar to IsoRank even though the matching algorithm defined here is different than the one in IsoRank.    

\subsection{Example-1: Alignment of Network Pairs}
\begin{figure*}
\centering
\subfigure[Probabilities for matching of the nodes of $G_2$ and $G_3$]{
\label{fig:graphsA}
\includegraphics[width=3in]{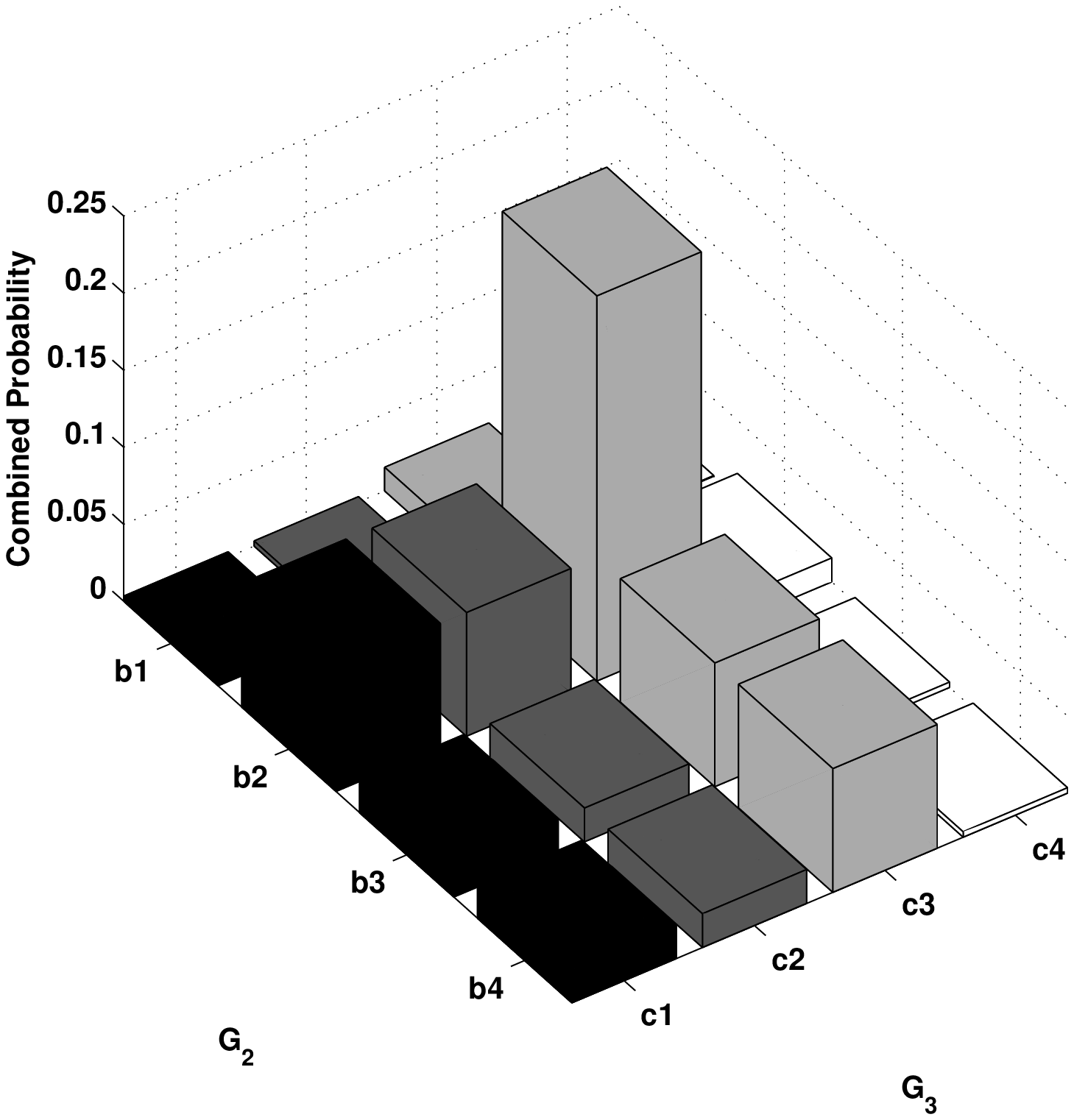}}
\subfigure[ Probabilities for matching of the nodes of $G_2$ and $G_3$ when the approximate Hermitian matrix is used.]{
\label{fig:graphsB}
\includegraphics[width=3in]{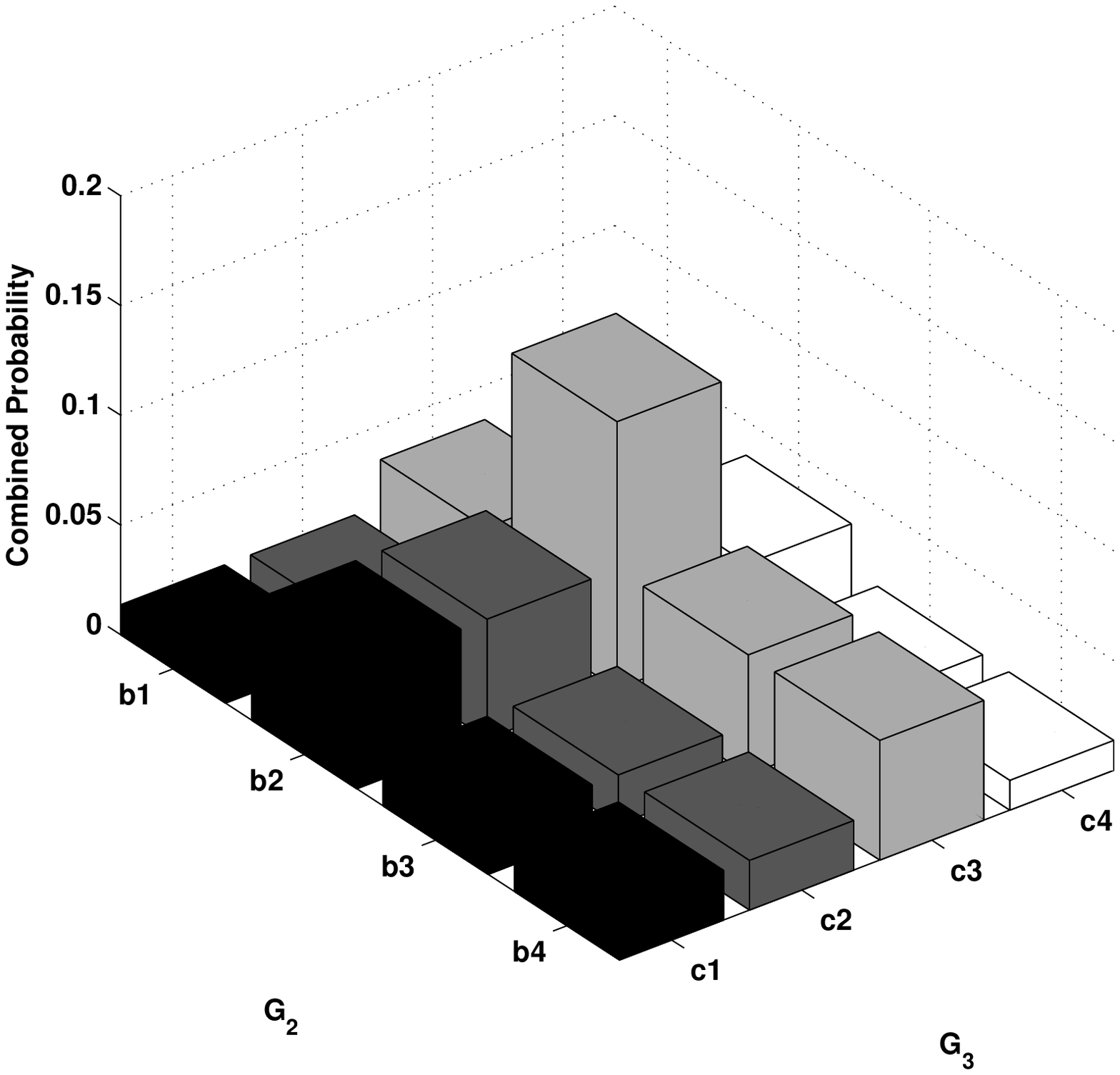}}
 \subfigure[Probabilities for matching of the nodes of $G_1$ and $G_2$]{
 \label{fig:graphsC}
\includegraphics[width=3in]{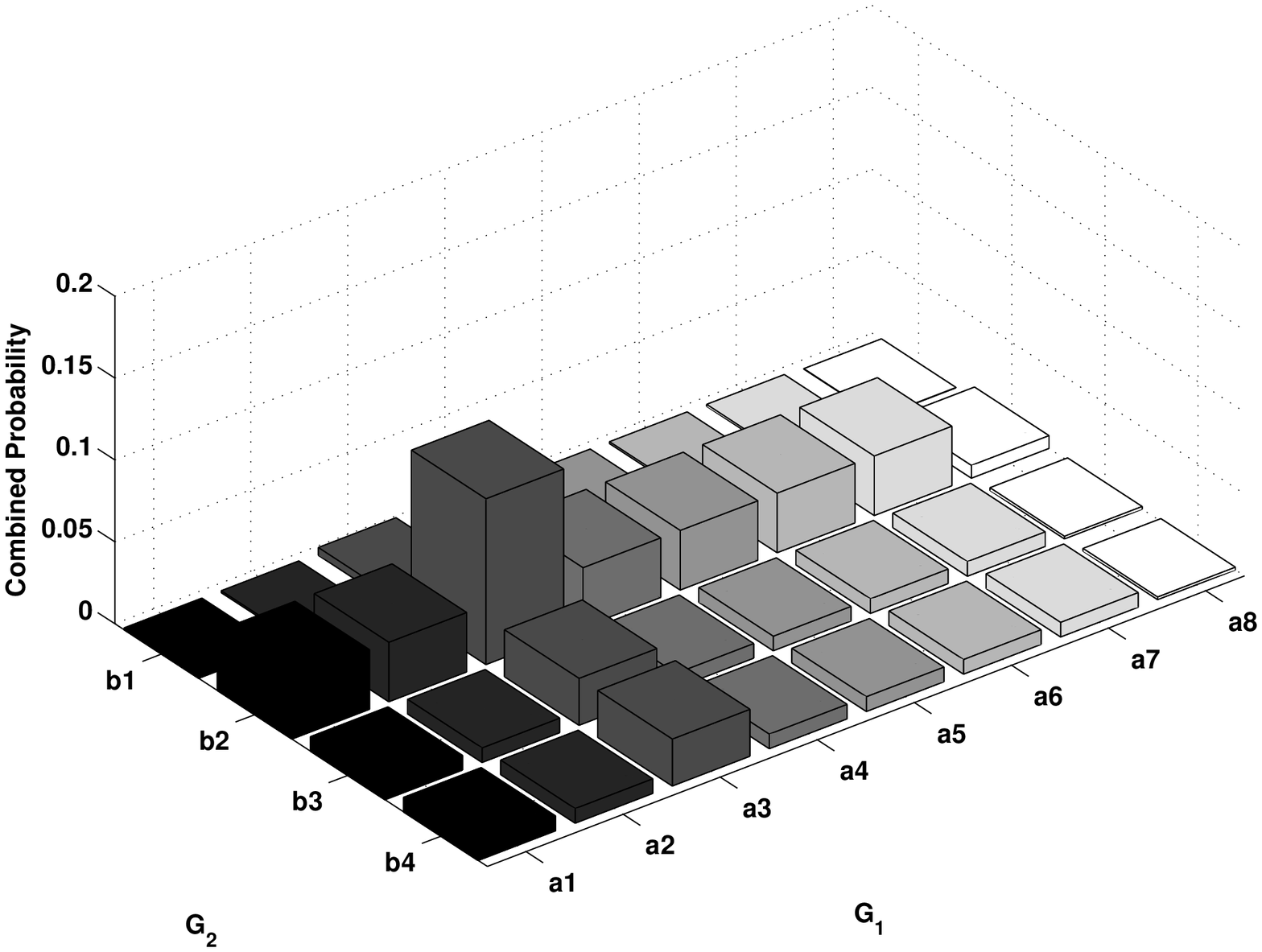}}
 \subfigure[Probabilities for matching of the nodes of $G_1$ and $G_2$ when the approximate Hermitian matrix is used.]{
\label{fig:graphsD}
\includegraphics[width=3in]{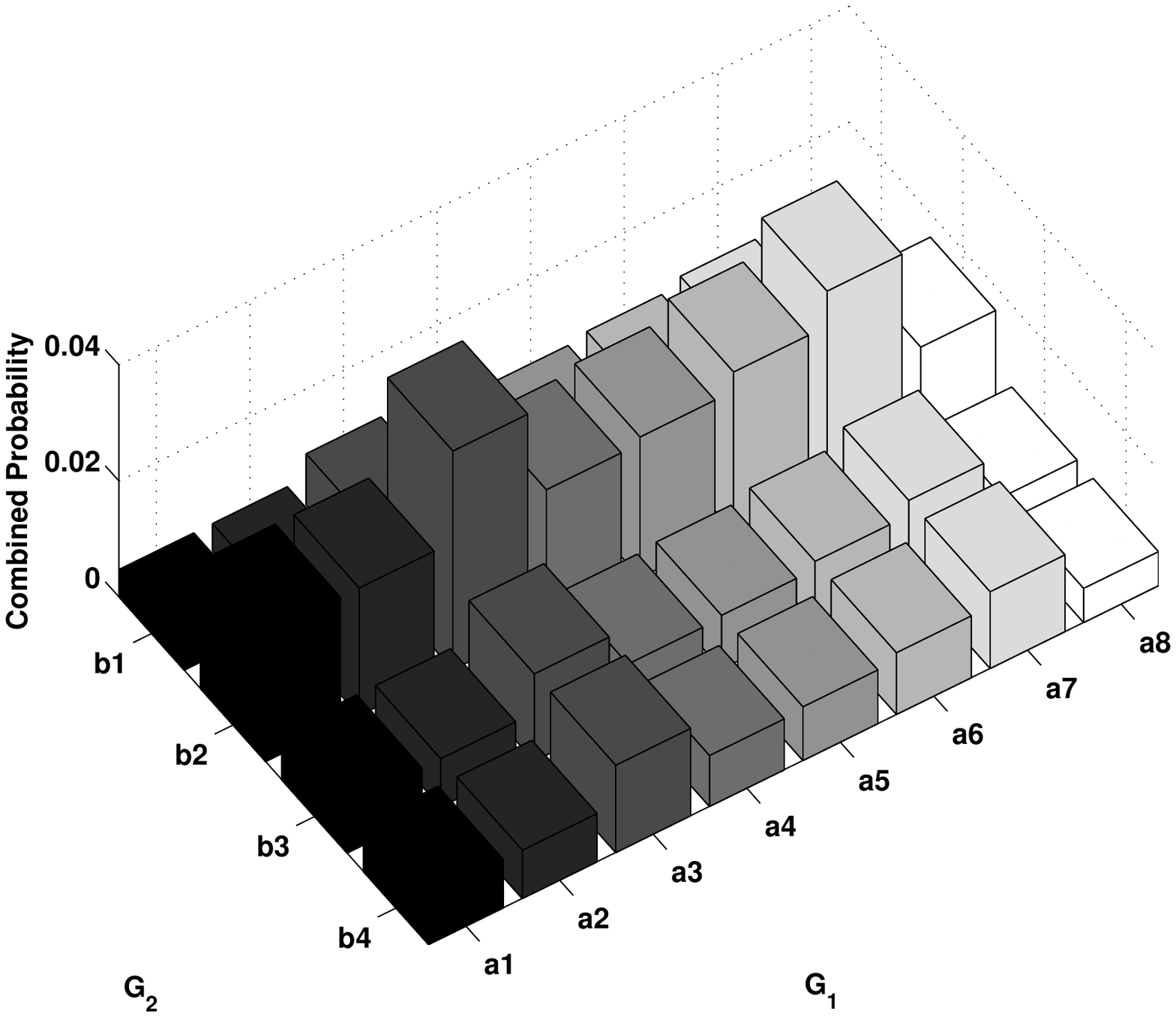}}
\caption{\label{fig:graphs1} Expected probability outcome for the alignment of pairs of the networks}
\end{figure*}

In the case of the alignment of two networks $G_1$-$G_2$, the possible  probability outcomes are shown in Fig.\ref{fig:graphsC}; where  it is first assumed that  $\ket{reg_{G_1}}=a_i$,  then probabilities in the collapsed state are found for  each $bj$.  
Note that  since the collapsed state is not normalized,  it  represents conditional probabilities: e.g., the probability of  $\ket{reg_{G_2}}=b_2$ when $\ket{reg_{G_1}}=a_1$. The same approach is also applied to $G_2$-$G_3$ and the probabilities are shown in Fig.\ref{fig:graphsA}. 
The matching algorithm applied to the outcomes  in Fig.\ref{fig:graphsA} and Fig.\ref{fig:graphsC} procures the exact alignments for the both pairs of the networks. 
To make the method understood well, we shall go through the alignment for $G_1$ and $G_2$ by assuming  PEA is run statistically enough times with the same setting: The  probabilities for $\ket{reg_{G_1}}$ in the final state of PEA are $[0.1176, 0.1176,  0.2647,0.1176, \dots,0.1176,
    0.0294]$ where we see $a_3$ with the highest probability. When $\ket{reg_{G_1}}=a_3$, the quantum state collapses to the following \underline{unnormalized state}: $[
   0.1213,
   0.3638,
   0.2425,
   0.2425]$ which produces to the following \underline{normalized probabilities}  $[ 0.0556,
    0.5000,
    0.2222,
    0.2222]$. From this state  we see $\ket{reg_{G_2}}=b_2$ with the highest probability. Hence, $a_3$ is matched to $b_2$. 
   In the next possible step, \ket{reg_{G_1}} can take the same probabilities for different states. In the measurement, we give the priority to the neighbours of  the matched nodes. When $\ket{reg_{G_1}}=a_1$, which is a neighbour of $a_3$, the quantum state collapses to $[0.0808
   0.2425,
   0.1617,
   0.1617]$ producing the normalized probabilities:  
   $[ 0.0556, 0.5000, 0.2222,  0.2222]$. Since $b_2$ is already matched, the second highest possible outcome  is either $b_3$ or $b_4$.  Therefore, $a_1$ is matched to either $b_3$ or $b_4$. In the final round, we match $a_2$ to the remaining one of $b_3$ and $b_4$. The same but \underline{unnormalized} probabilities are also shown in Fig.\ref{fig:graphsC}.

Fig.\ref{fig:graphsB} and Fig.\ref{fig:graphsD}  also show the probability outcomes for the alignment of the same networks: $G_1$-$G_2$ and $G_2$-$G_3$; however, in these figures, instead of $A$, approximate Hermitian matrices found by $H=1/2(A+A^\dagger)$ are used.  While the probabilities slightly differ in comparison to Fig.\ref{fig:graphsA} and Fig.\ref{fig:graphsC}, they still conclude with the same alignments.

\subsection{Example-2: Alignment of Multiple Networks}
\begin{figure*}[h]
\centering
\subfigure[Probabilities for matching of the nodes of $G_1$, $G_2$, and $G_3$.]{
\label{fig:graphsMN1}
\includegraphics[width=5in]{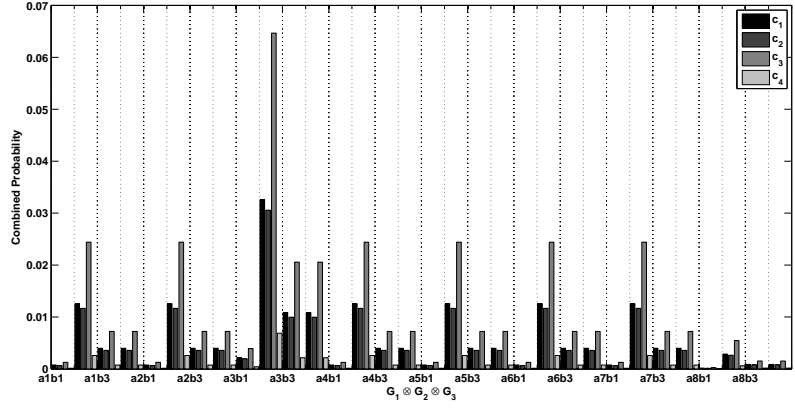}}
\subfigure[Probabilities for matching of the nodes of $G_1$, $G_2$, and $G_3$ when the approximate Hermitian matrix is used.]{
\label{fig:graphsMN2}
\includegraphics[width=5in]{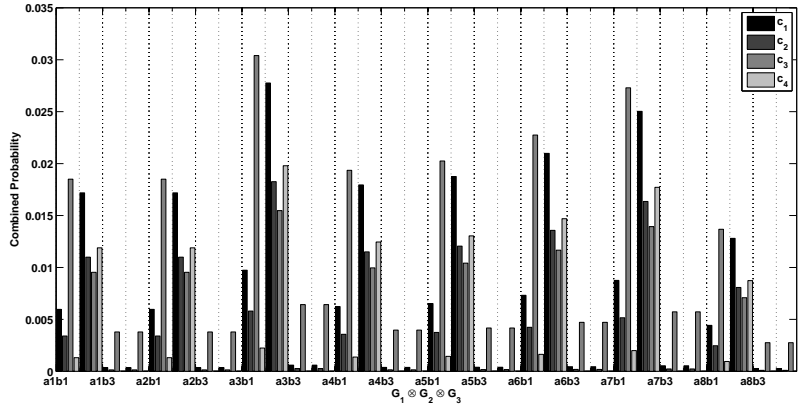}}
\caption{\label{fig:graphs2} Probabilities for matching of the nodes $G_1$, $G_2$, and $G_3$. While the x-axis represents the nodes, $a_ib_jc_k$; the y-axis is the  \underline{unnormalized} conditional probabilities: if $\ket{reg_{G_1}}$ is in $a_i$ state, then the probability to measure $\ket{reg_{G_2}}\ket{reg_{G_3}}$ in $b_jc_k$ state in the collapsed quantum state is shown. }
\end{figure*}

As an example for the multiple network alignment, we use $G_1$, $G_2$ and $G_3$ shown in Fig.\ref{fig:networks}  and follow a similar method to  the alignment of two networks: 
First, we assume $\ket{reg_{G_1}}=a_i$, then find the probability of seeing $\ket{reg_{G_2}reg_{G_2}}=b_jc_k$ in the unnormalized collapsed state which represents the conditional probabilities. 
The probabilities are shown in Fig.\ref{fig:graphsMN1}, where the exact matrix is used. Fig.\ref{fig:graphsMN1} represents the probabilities when the Hermitian matrix $H=1/2(A+A^\dagger)$ is used. 
Giving the priority to the neighbors of the first matched nodes as explained in Sec.\ref{sec:MatchingMN}, the exact same matching can be obtained from  both of the figures.

\section{The Overall Algorithmic Complexity}
\label{Sec:Complexity}
The computational complexity of a quantum algorithm is measured by the number of  fundamental quantum gates-e.g. one and two qubit gates-in the circuit which simulates the algorithm. The complexity of PEA is dominated by the complexity of the circuit implementing the given operator $A$. Implementation of $A$ order of $N$  requires $O(N^2)$ number of quantum gates \cite{Nielsen2010quantum}. However, it is proven that the efficient (polynomial time in the number of qubits) simulation  of a sparse operator  on quantum computers is possible  when the number of entries in $A$ is bounded polynomially in the number of qubits and the norm of the matrix is less than or equal to the degree of this polynomial \cite{Aharonov2003}. 
There have been also algorithms presented to simulate such sparse operators in polynomial time \cite{Berry2007sparse,Childs2011}. 
As a result, since the adjacency matrices of the PPI networks sparse, they can be simulated efficiently on quantum computers. 

In the implementation of the operator $A$,  quantum circuits for each graph can be generated separately since it is the Kronecker product of the column-wise normalized  adjacency matrices (normalized in the sense to make them stochastic): $A=A_1\otimes A_2 \otimes \dots \otimes A_m$, where $m$ represents the number of networks. Therefore, the total complexity of implementing $A$ can be defined as:
\begin{equation}
O\Big(poly\big(log(|V_1|)\big)+poly\big(log(|V_2|)\big) \dots poly\big(log(|V_m|)\big)\Big),
\end{equation}
or more concisely:
\begin{equation}
O\Big(m\times poly\big(log (|V_{max}|)\big)\Big),
\end{equation} 
where $|V_{max}|$ defines the maximum number of nodes in a graph. This indicates an exponentially faster  evaluation time for the implementation of $A$ than the implementations on classical computers. 
Note that if the adjacency matrices are not sparse, then the complexity for this part becomes $O\big(m|V_{max}|^2\big)$ which  still provides exponential speed-up in comparison to $O\big(|E|^m\big)$ classical complexity.
Moreover, in the case of the incorporation of additional data: i.e., using $A+B$ instead of $A$; the above complexity arguments hold by assuming $A$ and $B$ commute and $B$ is efficiently simulatable. 

The complexity of the matching part of the algorithm is related to the number of measurements applied to the system. If only one kind of measurement setting (measure a register alone and the rest together to draw the conditional probabilities) is used, then this part  requires polynomial time (polynomial by the number of qubits) because it is related to the number of qubits. 
On the other hand, if one also uses different combinations of the registers in the measurements and consider them together to match the nodes (This can be only useful when additional data is incorporated.), then the complexity to store the statistical results of the measurement outcomes and find matching from these outcomes may grow exponentially with the number of networks.  However, as mentioned in Sec.\ref{sec:MatchingMN}, if no additional data is used, then the system is separable and all possible measurement settings produce the same output. Therefore, only one measurement setting can be used.

\section{Conclusion}
In this paper, we have presented a quantum approach for the alignment of multiple networks by adapting  quantum phase estimation algorithm. 
In particular, we have showed that the principal eigenvector of  a stochastic matrix used in IsoRank algorithm for the alignment can be found exponentially more efficiently on quantum phase estimation algorithm. 
Since the final quantum state  representing the eigenvector in the phase estimation algorithm is not classically available,  adapting a conditional measurement scheme, we have also showed a matching algorithm to obtain the alignment result from this state. 
In addition, since the stochastic matrices are generally not Hermitian, we have also discussed how to approximate them for the simulation on quantum computers. Finally, we have used three simple networks and showed the numerical alignment results for them. While the approach discussed here follows mainly IsoRank algorithm, we believe it shall also pave the way for the applications of other spectral alignment methods on quantum computers.
\section{Acknowledgment}
This work is supported by TUBITAK under the project number 115E747.
\bibliography{manuscript2}

\begin{thebibliography}{21}%
\makeatletter
\providecommand \@ifxundefined [1]{%
 \@ifx{#1\undefined}
}%
\providecommand \@ifnum [1]{%
 \ifnum #1\expandafter \@firstoftwo
 \else \expandafter \@secondoftwo
 \fi
}%
\providecommand \@ifx [1]{%
 \ifx #1\expandafter \@firstoftwo
 \else \expandafter \@secondoftwo
 \fi
}%
\providecommand \natexlab [1]{#1}%
\providecommand \enquote  [1]{``#1''}%
\providecommand \bibnamefont  [1]{#1}%
\providecommand \bibfnamefont [1]{#1}%
\providecommand \citenamefont [1]{#1}%
\providecommand \href@noop [0]{\@secondoftwo}%
\providecommand \href [0]{\begingroup \@sanitize@url \@href}%
\providecommand \@href[1]{\@@startlink{#1}\@@href}%
\providecommand \@@href[1]{\endgroup#1\@@endlink}%
\providecommand \@sanitize@url [0]{\catcode `\\12\catcode `\$12\catcode
  `\&12\catcode `\#12\catcode `\^12\catcode `\_12\catcode `\%12\relax}%
\providecommand \@@startlink[1]{}%
\providecommand \@@endlink[0]{}%
\providecommand \url  [0]{\begingroup\@sanitize@url \@url }%
\providecommand \@url [1]{\endgroup\@href {#1}{\urlprefix }}%
\providecommand \urlprefix  [0]{URL }%
\providecommand \Eprint [0]{\href }%
\providecommand \doibase [0]{http://dx.doi.org/}%
\providecommand \selectlanguage [0]{\@gobble}%
\providecommand \bibinfo  [0]{\@secondoftwo}%
\providecommand \bibfield  [0]{\@secondoftwo}%
\providecommand \translation [1]{[#1]}%
\providecommand \BibitemOpen [0]{}%
\providecommand \bibitemStop [0]{}%
\providecommand \bibitemNoStop [0]{.\EOS\space}%
\providecommand \EOS [0]{\spacefactor3000\relax}%
\providecommand \BibitemShut  [1]{\csname bibitem#1\endcsname}%
\let\auto@bib@innerbib\@empty
\bibitem [{\citenamefont {Klau}(2009)}]{KLau2009}%
  \BibitemOpen
  \bibfield  {author} {\bibinfo {author} {\bibfnamefont {G.}~\bibnamefont
  {Klau}},\ }\href {\doibase 10.1186/1471-2105-10-S1-S59} {\bibfield  {journal}
  {\bibinfo  {journal} {BMC Bioinformatics}\ }\textbf {\bibinfo {volume}
  {10}},\ \bibinfo {pages} {S59} (\bibinfo {year} {2009})}\BibitemShut
  {NoStop}%
\bibitem [{\citenamefont {Kelley}\ \emph {et~al.}(2004)\citenamefont {Kelley},
  \citenamefont {Yuan}, \citenamefont {Lewitter}, \citenamefont {Sharan},
  \citenamefont {Stockwell},\ and\ \citenamefont {Ideker}}]{Kelley2004}%
  \BibitemOpen
  \bibfield  {author} {\bibinfo {author} {\bibfnamefont {B.~P.}\ \bibnamefont
  {Kelley}}, \bibinfo {author} {\bibfnamefont {B.}~\bibnamefont {Yuan}},
  \bibinfo {author} {\bibfnamefont {F.}~\bibnamefont {Lewitter}}, \bibinfo
  {author} {\bibfnamefont {R.}~\bibnamefont {Sharan}}, \bibinfo {author}
  {\bibfnamefont {B.~R.}\ \bibnamefont {Stockwell}}, \ and\ \bibinfo {author}
  {\bibfnamefont {T.}~\bibnamefont {Ideker}},\ }\href {\doibase
  10.1093/nar/gkh411} {\bibfield  {journal} {\bibinfo  {journal} {Nucleic Acids
  Research}\ }\textbf {\bibinfo {volume} {32}},\ \bibinfo {pages} {W83}
  (\bibinfo {year} {2004})}\BibitemShut {NoStop}%
\bibitem [{\citenamefont {Koyut{\"u}rk}\ \emph {et~al.}(2006)\citenamefont
  {Koyut{\"u}rk}, \citenamefont {Kim}, \citenamefont {Topkara}, \citenamefont
  {Subramaniam}, \citenamefont {Szpankowski},\ and\ \citenamefont
  {Grama}}]{Koyuturk2006pairwise}%
  \BibitemOpen
  \bibfield  {author} {\bibinfo {author} {\bibfnamefont {M.}~\bibnamefont
  {Koyut{\"u}rk}}, \bibinfo {author} {\bibfnamefont {Y.}~\bibnamefont {Kim}},
  \bibinfo {author} {\bibfnamefont {U.}~\bibnamefont {Topkara}}, \bibinfo
  {author} {\bibfnamefont {S.}~\bibnamefont {Subramaniam}}, \bibinfo {author}
  {\bibfnamefont {W.}~\bibnamefont {Szpankowski}}, \ and\ \bibinfo {author}
  {\bibfnamefont {A.}~\bibnamefont {Grama}},\ }\href@noop {} {\bibfield
  {journal} {\bibinfo  {journal} {Journal of Computational Biology}\ }\textbf
  {\bibinfo {volume} {13}},\ \bibinfo {pages} {182} (\bibinfo {year}
  {2006})}\BibitemShut {NoStop}%
\bibitem [{\citenamefont {Li}\ \emph {et~al.}(2007)\citenamefont {Li},
  \citenamefont {Zhang}, \citenamefont {Wang}, \citenamefont {Zhang},\ and\
  \citenamefont {Chen}}]{Li2007alignment}%
  \BibitemOpen
  \bibfield  {author} {\bibinfo {author} {\bibfnamefont {Z.}~\bibnamefont
  {Li}}, \bibinfo {author} {\bibfnamefont {S.}~\bibnamefont {Zhang}}, \bibinfo
  {author} {\bibfnamefont {Y.}~\bibnamefont {Wang}}, \bibinfo {author}
  {\bibfnamefont {X.-S.}\ \bibnamefont {Zhang}}, \ and\ \bibinfo {author}
  {\bibfnamefont {L.}~\bibnamefont {Chen}},\ }\href@noop {} {\bibfield
  {journal} {\bibinfo  {journal} {Bioinformatics}\ }\textbf {\bibinfo {volume}
  {23}},\ \bibinfo {pages} {1631} (\bibinfo {year} {2007})}\BibitemShut
  {NoStop}%
\bibitem [{\citenamefont {Zaslavskiy}\ \emph {et~al.}(2009)\citenamefont
  {Zaslavskiy}, \citenamefont {Bach},\ and\ \citenamefont
  {Vert}}]{Zaslavskiy15062009}%
  \BibitemOpen
  \bibfield  {author} {\bibinfo {author} {\bibfnamefont {M.}~\bibnamefont
  {Zaslavskiy}}, \bibinfo {author} {\bibfnamefont {F.}~\bibnamefont {Bach}}, \
  and\ \bibinfo {author} {\bibfnamefont {J.-P.}\ \bibnamefont {Vert}},\ }\href
  {\doibase 10.1093/bioinformatics/btp196} {\bibfield  {journal} {\bibinfo
  {journal} {Bioinformatics}\ }\textbf {\bibinfo {volume} {25}},\ \bibinfo
  {pages} {i259} (\bibinfo {year} {2009})}\BibitemShut {NoStop}%
\bibitem [{\citenamefont {Clark}\ and\ \citenamefont
  {Kalita}(2014)}]{Clark2014comparison}%
  \BibitemOpen
  \bibfield  {author} {\bibinfo {author} {\bibfnamefont {C.}~\bibnamefont
  {Clark}}\ and\ \bibinfo {author} {\bibfnamefont {J.}~\bibnamefont {Kalita}},\
  }\href@noop {} {\bibfield  {journal} {\bibinfo  {journal} {Bioinformatics}\
  }\textbf {\bibinfo {volume} {30}},\ \bibinfo {pages} {2351} (\bibinfo {year}
  {2014})}\BibitemShut {NoStop}%
\bibitem [{\citenamefont {Page}\ \emph {et~al.}(1999)\citenamefont {Page},
  \citenamefont {Brin}, \citenamefont {Motwani},\ and\ \citenamefont
  {Winograd}}]{PageRank}%
  \BibitemOpen
  \bibfield  {author} {\bibinfo {author} {\bibfnamefont {L.}~\bibnamefont
  {Page}}, \bibinfo {author} {\bibfnamefont {S.}~\bibnamefont {Brin}}, \bibinfo
  {author} {\bibfnamefont {R.}~\bibnamefont {Motwani}}, \ and\ \bibinfo
  {author} {\bibfnamefont {T.}~\bibnamefont {Winograd}},\ }\href@noop {} {\emph
  {\bibinfo {title} {The PageRank Citation Ranking: Bringing Order to the
  Web.}}},\ \bibinfo {type} {Technical Report}\ \bibinfo {number} {1999-66}\
  (\bibinfo  {institution} {Stanford InfoLab},\ \bibinfo {year} {1999})\
  \bibinfo {note} {previous number = SIDL-WP-1999-0120}\BibitemShut {NoStop}%
\bibitem [{\citenamefont {Singh}\ \emph {et~al.}(2007)\citenamefont {Singh},
  \citenamefont {Xu},\ and\ \citenamefont {Berger}}]{Singh2007}%
  \BibitemOpen
  \bibfield  {author} {\bibinfo {author} {\bibfnamefont {R.}~\bibnamefont
  {Singh}}, \bibinfo {author} {\bibfnamefont {J.}~\bibnamefont {Xu}}, \ and\
  \bibinfo {author} {\bibfnamefont {B.}~\bibnamefont {Berger}},\ }in\ \href
  {\doibase 10.1007/978-3-540-71681-5_2} {\emph {\bibinfo {booktitle} {Research
  in Computational Molecular Biology}}},\ \bibinfo {series} {Lecture Notes in
  Computer Science}, Vol.\ \bibinfo {volume} {4453},\ \bibinfo {editor} {edited
  by\ \bibinfo {editor} {\bibfnamefont {T.}~\bibnamefont {Speed}}\ and\
  \bibinfo {editor} {\bibfnamefont {H.}~\bibnamefont {Huang}}}\ (\bibinfo
  {publisher} {Springer Berlin Heidelberg},\ \bibinfo {year} {2007})\ pp.\
  \bibinfo {pages} {16--31}\BibitemShut {NoStop}%
\bibitem [{\citenamefont {Liao}\ \emph {et~al.}(2009)\citenamefont {Liao},
  \citenamefont {Lu}, \citenamefont {Baym}, \citenamefont {0001},\ and\
  \citenamefont {Berger}}]{IsorankN2009}%
  \BibitemOpen
  \bibfield  {author} {\bibinfo {author} {\bibfnamefont {C.-S.}\ \bibnamefont
  {Liao}}, \bibinfo {author} {\bibfnamefont {K.}~\bibnamefont {Lu}}, \bibinfo
  {author} {\bibfnamefont {M.}~\bibnamefont {Baym}}, \bibinfo {author}
  {\bibfnamefont {R.~S.}\ \bibnamefont {0001}}, \ and\ \bibinfo {author}
  {\bibfnamefont {B.}~\bibnamefont {Berger}},\ }\href@noop {} {\bibfield
  {journal} {\bibinfo  {journal} {Bioinformatics}\ }\textbf {\bibinfo {volume}
  {25}} (\bibinfo {year} {2009})}\BibitemShut {NoStop}%
\bibitem [{\citenamefont {Bayati}\ \emph {et~al.}(2009)\citenamefont {Bayati},
  \citenamefont {Gerritsen}, \citenamefont {Gleich}, \citenamefont {Saberi},\
  and\ \citenamefont {Wang}}]{Wang2009}%
  \BibitemOpen
  \bibfield  {author} {\bibinfo {author} {\bibfnamefont {M.}~\bibnamefont
  {Bayati}}, \bibinfo {author} {\bibfnamefont {M.}~\bibnamefont {Gerritsen}},
  \bibinfo {author} {\bibfnamefont {D.}~\bibnamefont {Gleich}}, \bibinfo
  {author} {\bibfnamefont {A.}~\bibnamefont {Saberi}}, \ and\ \bibinfo {author}
  {\bibfnamefont {Y.}~\bibnamefont {Wang}},\ }in\ \href@noop {} {\emph
  {\bibinfo {booktitle} {Data Mining, 2009. ICDM '09. Ninth IEEE International
  Conference on}}}\ (\bibinfo {year} {2009})\ pp.\ \bibinfo {pages}
  {705--710}\BibitemShut {NoStop}%
\bibitem [{\citenamefont {Nir}\ \emph {et~al.}(2008)\citenamefont {Nir},
  \citenamefont {Roded},\ and\ \citenamefont {William~Stafford}}]{Noble2008}%
  \BibitemOpen
  \bibfield  {author} {\bibinfo {author} {\bibfnamefont {Y.}~\bibnamefont
  {Nir}}, \bibinfo {author} {\bibfnamefont {S.}~\bibnamefont {Roded}}, \ and\
  \bibinfo {author} {\bibfnamefont {N.}~\bibnamefont {William~Stafford}},\
  }\href@noop {} {\bibfield  {journal} {\bibinfo  {journal} {Bioinformatics}\
  }\textbf {\bibinfo {volume} {24}},\ \bibinfo {pages} {i200} (\bibinfo {year}
  {2008})}\BibitemShut {NoStop}%
\bibitem [{\citenamefont {Daskin}\ \emph {et~al.}(2014)\citenamefont {Daskin},
  \citenamefont {Grama},\ and\ \citenamefont {Kais}}]{Daskin2014mna}%
  \BibitemOpen
  \bibfield  {author} {\bibinfo {author} {\bibfnamefont {A.}~\bibnamefont
  {Daskin}}, \bibinfo {author} {\bibfnamefont {A.}~\bibnamefont {Grama}}, \
  and\ \bibinfo {author} {\bibfnamefont {S.}~\bibnamefont {Kais}},\ }\href@noop
  {} {\bibfield  {journal} {\bibinfo  {journal} {Quantum Information
  Processing}\ }\textbf {\bibinfo {volume} {13}},\ \bibinfo {pages} {2653}
  (\bibinfo {year} {2014})}\BibitemShut {NoStop}%
\bibitem [{\citenamefont {Cook}(1971)}]{Cook1971}%
  \BibitemOpen
  \bibfield  {author} {\bibinfo {author} {\bibfnamefont {S.~A.}\ \bibnamefont
  {Cook}},\ }in\ \href@noop {} {\emph {\bibinfo {booktitle} {Proceedings of the
  third annual ACM symposium on Theory of computing}}}\ (\bibinfo
  {organization} {ACM},\ \bibinfo {year} {1971})\ pp.\ \bibinfo {pages}
  {151--158}\BibitemShut {NoStop}%
\bibitem [{\citenamefont {Sharan}\ and\ \citenamefont
  {Ideker}(2006)}]{Sharan2006modeling}%
  \BibitemOpen
  \bibfield  {author} {\bibinfo {author} {\bibfnamefont {R.}~\bibnamefont
  {Sharan}}\ and\ \bibinfo {author} {\bibfnamefont {T.}~\bibnamefont
  {Ideker}},\ }\href@noop {} {\bibfield  {journal} {\bibinfo  {journal} {Nature
  biotechnology}\ }\textbf {\bibinfo {volume} {24}},\ \bibinfo {pages} {427}
  (\bibinfo {year} {2006})}\BibitemShut {NoStop}%
\bibitem [{\citenamefont {Nielsen}\ and\ \citenamefont
  {Chuang}(2010)}]{Nielsen2010quantum}%
  \BibitemOpen
  \bibfield  {author} {\bibinfo {author} {\bibfnamefont {M.~A.}\ \bibnamefont
  {Nielsen}}\ and\ \bibinfo {author} {\bibfnamefont {I.~L.}\ \bibnamefont
  {Chuang}},\ }\href@noop {} {\emph {\bibinfo {title} {Quantum computation and
  quantum information}}}\ (\bibinfo  {publisher} {Cambridge university press},\
  \bibinfo {year} {2010})\BibitemShut {NoStop}%
\bibitem [{\citenamefont {Kitaev}(1995)}]{Kitaev1995}%
  \BibitemOpen
  \bibfield  {author} {\bibinfo {author} {\bibfnamefont {A.}~\bibnamefont
  {Kitaev}},\ }\href {http://arxiv.org/abs/quant-ph/9511026} {\bibfield
  {journal} {\bibinfo  {journal} {arXiv:quant-ph/9511026}\ } (\bibinfo {year}
  {1995})}\BibitemShut {NoStop}%
\bibitem [{\citenamefont {Keller}(1975)}]{ClosestHermitian1975}%
  \BibitemOpen
  \bibfield  {author} {\bibinfo {author} {\bibfnamefont {J.~B.}\ \bibnamefont
  {Keller}},\ }\href@noop {} {\bibfield  {journal} {\bibinfo  {journal}
  {Mathematics Magazine}\ }\textbf {\bibinfo {volume} {48}},\ \bibinfo {pages}
  {pp. 192} (\bibinfo {year} {1975})}\BibitemShut {NoStop}%
\bibitem [{\citenamefont {Berman}\ and\ \citenamefont
  {Plemmons}(1994)}]{Berman1994}%
  \BibitemOpen
  \bibfield  {author} {\bibinfo {author} {\bibfnamefont {A.}~\bibnamefont
  {Berman}}\ and\ \bibinfo {author} {\bibfnamefont {R.~J.}\ \bibnamefont
  {Plemmons}},\ }\enquote {\bibinfo {title} {4. symmetric nonnegative
  matrices},}\ in\ \href@noop {} {\emph {\bibinfo {booktitle} {Nonnegative
  Matrices in the Mathematical Sciences}}}\ (\bibinfo {year} {1994})\
  Chap.~\bibinfo {chapter} {4}, pp.\ \bibinfo {pages} {87--111}\BibitemShut
  {NoStop}%
\bibitem [{\citenamefont {Berry}\ \emph {et~al.}(2007)\citenamefont {Berry},
  \citenamefont {Ahokas}, \citenamefont {Cleve},\ and\ \citenamefont
  {Sanders}}]{Berry2007sparse}%
  \BibitemOpen
  \bibfield  {author} {\bibinfo {author} {\bibfnamefont {D.}~\bibnamefont
  {Berry}}, \bibinfo {author} {\bibfnamefont {G.}~\bibnamefont {Ahokas}},
  \bibinfo {author} {\bibfnamefont {R.}~\bibnamefont {Cleve}}, \ and\ \bibinfo
  {author} {\bibfnamefont {B.}~\bibnamefont {Sanders}},\ }\href {\doibase
  10.1007/s00220-006-0150-x} {\bibfield  {journal} {\bibinfo  {journal}
  {Communications in Mathematical Physics}\ }\textbf {\bibinfo {volume}
  {270}},\ \bibinfo {pages} {359} (\bibinfo {year} {2007})}\BibitemShut
  {NoStop}%
\bibitem [{\citenamefont {Childs}\ and\ \citenamefont
  {Kothari}(2011)}]{Childs2011}%
  \BibitemOpen
  \bibfield  {author} {\bibinfo {author} {\bibfnamefont {A.~M.}\ \bibnamefont
  {Childs}}\ and\ \bibinfo {author} {\bibfnamefont {R.}~\bibnamefont
  {Kothari}},\ }in\ \href@noop {} {\emph {\bibinfo {booktitle} {Theory of
  Quantum Computation, Communication, and Cryptography}}}\ (\bibinfo
  {publisher} {Springer},\ \bibinfo {year} {2011})\ pp.\ \bibinfo {pages}
  {94--103}\BibitemShut {NoStop}%
\bibitem [{\citenamefont {Aharonov}\ and\ \citenamefont
  {Ta-Shma}(2003)}]{Aharonov2003}%
  \BibitemOpen
  \bibfield  {author} {\bibinfo {author} {\bibfnamefont {D.}~\bibnamefont
  {Aharonov}}\ and\ \bibinfo {author} {\bibfnamefont {A.}~\bibnamefont
  {Ta-Shma}},\ }in\ \href@noop {} {\emph {\bibinfo {booktitle} {Proceedings of
  the thirty-fifth annual ACM symposium on Theory of computing}}},\ \bibinfo
  {series and number} {STOC '03}\ (\bibinfo  {publisher} {ACM},\ \bibinfo
  {address} {New York, NY, USA},\ \bibinfo {year} {2003})\ pp.\ \bibinfo
  {pages} {20--29}\BibitemShut {NoStop}%
\end{thebibliography}%

\end{document}